\documentclass[fleqn,usenatbib]{mnras}

\usepackage{newtxtext,newtxmath}
\usepackage[T1]{fontenc}

\DeclareRobustCommand{\VAN}[3]{#2}
\let\VANthebibliography\thebibliography
\def\thebibliography{\DeclareRobustCommand{\VAN}[3]{##3}\VANthebibliography}

\usepackage{graphicx}	% Including figure files
\usepackage{amsmath}	% Advanced maths commands
\title[BAL Masking]{The impact and mitigation of broad absorption line quasars in Lyman$-\alpha$ forest correlations}

\author[L. Ennesser et al.]{
Lauren Ennesser,$^{1,2}$\thanks{E-mail: ennesser.1@osu.edu}
Paul Martini,$^{2,3,4}$ Andreu Font-Ribera,$^{5,6}$ Ignasi P\'{e}rez-R\`{a}fols$^{7}$
\\
$^{1}$Department of Physics, The Ohio State University, Columbus, Ohio 43210, USA \\
$^{2}$Center of Cosmology and Astro-Particle Physics, The Ohio State University, Columbus, Ohio, 43210, USA \\
$^{3}$Department of Astronomy, The Ohio State University, Columbus, Ohio 43210, USA \\
$^{4}$Radcliffe Institute for Advanced Study, Harvard University, Cambridge, MA 02138, USA \\
$^{5}$Institut de Física d’Altes Energies, The Barcelona Institute of Science and Technology, Campus UAB, 08193 Bellaterra (Barcelona), Spain \\ 
$^{6}$Department of Physics and Astronomy, University College London, Gower Street, London WC1E 6BT, UK. \\
$^{7}$Sorbonne Université, Université Paris Diderot, CNRS/IN2P3, Laboratoire de Physique Nucléaire et de Hautes Energies, \\ LPNHE, 4 Place Jussieu, F-75252 Paris, France \\   
}

\date{Accepted XXX. Received YYY; in original form ZZZ}

\pubyear{2021}

\begin{document}
\label{firstpage}
\pagerange{\pageref{firstpage}--\pageref{lastpage}}
\maketitle

\begin{abstract}
Correlations in and with the flux transmission of the Lyman$-\alpha$ (Ly$\alpha$) forest in the spectra of high-redshift quasars are powerful cosmological tools, yet these measurements can be compromised if the intrinsic quasar continuum is significantly uncertain. One particularly problematic case is broad absorption line (BAL) quasars, which exhibit blueshifted absorption associated with many spectral features that are consistent with outflows of up to $\sim0.1c$. As these absorption features can both fall in the forest region and be difficult to distinguish from Ly$\alpha$ absorption, cosmological analyses eliminate the $\sim12 - 16$\% of quasars that exhibit BALs. In this paper we explore an alternate approach that includes BALs in the Ly$\alpha$ auto correlation function, with the exception of the expected locations of the BAL absorption troughs. This procedure returns over 95\% of the pathlength that is lost by the exclusion of BALs, as well as increases the density of sightlines. We show that including BAL quasars reduces the fractional uncertainty in the covariance matrix and correlation function by 12\% and does not significantly change the shape of the correlation function relative to analyses that exclude BAL quasars. We also evaluate different definitions of BALs, masking strategies, and potential differences in the quasar continuum in the forest region for BALs with different amounts of absorption. 
\end{abstract}


\section{Introduction}

% 1. Cosmic acceleration

The physical reason for the accelerating expansion of the universe is one of the greatest mysteries of modern physics. Measurements from Type Ia supernoave (SNe) provided the first clear evidence for this acceleration \citep{riess98,perlmutter99}. Subsequent measurements of the Baryon Acoustic Oscillation (BAO) scale \citep{eisenstein05,cole05} provided important, further support of the acceleration of cosmic expansion with independent distance measurements relative to the sound horizon, a physical scale that is exquisitely measured from CMB anisotropies \citep[e.g.][]{planck16}. \citet{Weinberg_2013} provides a detailed review of observational probes of cosmic acceleration.

The simplest model that fits the expansion history is the Cold Dark Matter model with a cosmological constant ($\Lambda$CDM) as expressed in Einstein's field equations. The cosmological constant acts like a negative pressure associated with the vacuum of space and is most commonly described as 'dark energy.' Yet the term dark energy does not provide a physical explanation of the acceleration, nor do present data rule out other models for cosmic acceleration, such as time variation in the dark energy component. The search for a physical explanation for cosmic acceleration is consequently a major question in fundamental physics, and has motivated many dark energy experiments. These experiments are commonly compared based on their forecast joint uncertainty on the dark energy equation of state parameter $w$ and the time variation $w'$, a metric proposed by the Dark Energy Task Force \citep[DETF][]{albrecht06}. The search for time variation is particularly appealing because it would provide strong support for certain physical models.

% 2. BAO at low redshift

At redshifts less than $z \sim 2$ the expansion history beyond the local universe is measured with both Type Ia SNe \citep{jha06,frieman08,guy10} and the BAO scale \citep{percival07,ross15}. The BAO measurements include large numbers of redshifts of individual galaxies and quasars from the 2dF Galaxy Redshift Survey \citep{colless01,percival01}, the Sloan Digital Sky Survey \citep[SDSS][]{york00}, and the 6dF Galaxy Redshift Survey \citep{beutler11}. The SDSS project has been extended into new generations multiple times, including dedicated cosmology programs called the Baryon Oscillation Spectroscopic Survey \citep[BOSS,][]{dawson13} as part of SDSS-III \citep{eisenstein11} and the more recent extended-BOSS \citep[eBOSS,][]{dawson16} as part of SDSS-IV \citep{blanton17}. The final results from eBOSS provide extremely strong support for dark energy, even in a model that allows for free curvature and a time-evolving equation of state for dark energy \citep{2021PhRvD.103h3533A}.

% 3. Use of the LyA at high redshift

At progressively higher redshifts, quasars replace galaxies as the highest number density sources at readily accessible flux limits. Above a redshift of $z\sim2$, quasars become especially valuable because the Lyman-$\alpha$ forest becomes accessible with ground-based observations. This is important because spectra of the absorption in the Ly$\alpha$ forest contains information about the distribution of matter (specifically neutral hydrogen) in the intergalactic medium (IGM) along the line of sight to the quasar, and not just at the position of the quasar \citep[e.g.][]{Slosar_2011,weinberg99}. The large, comoving pathlength sampled by Ly$\alpha$ forest absorption makes each quasar statistically more valuable than a discrete tracer, even though the Ly$\alpha$ absorption originates from lower density, less biased fluctuations in the matter power spectrum \citep{mcdonald03,white03}.

Large spectroscopic samples of quasars from BOSS and eBOSS were used to first measure the BAO signal in the auto-correlation function of the forest \citep{Busca_2013,Slosar_2013,kirkby13} and then the cross-correlation between the forest and quasars \citep{font14}. The continued increase in sample size led to progressively tighter constraints on the BAO scale, and culminated in the best measurement to date of $D_{H}/r_{d} = 8.99\pm 0.19$ and $D_{M}/r_{d}=37.5\pm 1.1$ at an effective redshift of $z = 2.33$ \citep{du_Mas_des_Bourboux_2020}, using data from the sixteenth and final eBOSS data release \citep[DR16][]{ahumada20}. This most recent measurement calculated the autocorrelation function with 210005 quasars at $z > 2.10$ and the cross-correlation function with 341468 quasars at $z > 1.77$.

% 4. Exclusion of BALs

Yet the impressive sample sizes used in these studies do not include all of the quasars above these redshifts that were observed. The main astrophysical reason some quasars are not included is that they show evidence of broad absorption line (BAL) troughs. The quasar subclass called BAL quasars were first defined by \citet{Weymann_1991} as quasars with broad absorption troughs blueshifted by at least $2000\,{\rm km}\,{\rm s}^{-1}$ and velocity widths of at least $2000\,{\rm km}\,{\rm s}^{-1}$. BAL troughs are most often observed associated with the \ion{C}{iv} spectral line, but are problematic for Ly$\alpha$ forest studies because they are also often associated with other spectral lines that may extend into the wavelength used for the forest analysis, such as \ion{N}{v} and Ly$\alpha$, and other lines that are in the forest region, such as \ion{P}{v} and \ion{S}{iv}. In early work by \citet{Slosar_2013}, BALs were included with the exception of the region in the immediate vicinity of Ly$\alpha$ and \ion{N}{v}. Yet because of concerns about other, weaker absorption features in the forest region, BALs were completely excluded from subsequent SDSS Ly$\alpha$ correlation studies.

BAL quasars typically comprise $10-30$\% of all quasars that employ UV and visible wavelength selection criteria \citep{foltz90,trump06} and most samples were identified via visual inspection. The largest such effort was the visual inspection of the 297,301 quasars in the SDSS Data Release 12 (DR12) quasar catalog by \citet{Paris_2017}, which identified 29,580 BALs, or $13$\% of the catalog above a redshift of $1.57$. The larger quasar samples in the DR14 \citep{Paris_2018} and DR16 \citep{lyke20} quasar catalogs were not all subject to visual inspection. Instead, BALs were identified in those catalogs with a combination of machine-learning methods \citep{Guo_2019} and other algorithms \citep{Paris_2018,lyke20}. Nevertheless, the BAL quasars have been excluded by all of the cosmological studies to a greater or lesser extent, depending on the threshold used to identify or classify BALs, and the result is a commensurate decrease in the available pathlength for the correlation studies.
%DR12 has 224844 objects with z_vi > 1.57

% 5. Summary of this paper

In this paper we study strategies to retain BAL quasars in Ly$\alpha$ forest analysis and quantify their impact on the correlation function measurements with data from the fourteenth data release of SDSS \citep[DR14,][]{abolfathi18}. The basic scheme is that we measure the velocity range impacted by the absorbing material from the blueshifted absorption troughs associated with the \ion{C}{iv} feature. This is a straightforward and conservative way to identify BALs, both because the quasar continuum is otherwise fairly featureless on the blueward side of \ion{C}{iv}, and because essentially all BAL quasars exhibit absorption associated with \ion{C}{iv} (although the converse is not true). We then mask the corresponding wavelengths that would have absorption troughs if the BALs were also associated with other spectral features and exclude those wavelengths from the forest analysis. These spectral features are known from both space-based spectra of low-redshift BALs \citep{arav98,leighly09} that are less affected by forest absorption and stacked spectra of SDSS BAL quasars \citep{Mas_Ribas_2019, hamann19}. 
%While this approach may exclude more pathlength than is actually affected, in practice there is little difference between our conservative approach and not excluding any pathlength.

In \S\ref{sec:bals} we review the two common ways to characterize BALs in the literature, summarize the catalogs we use for this study, and provide a detailed description of our masking procedure. In \S\ref{sec:cont} we describe various subsamples of BALs and explore differences in their continuum shape and the impact of masking. This investigation is important because current Ly$\alpha$ forest analysis codes fit a mean quasar continuum model to all quasars with only a zeropoint and slope term to account for the intrinsic diversity of quasar continuum shapes. We then compute the auto correlation function for these subsamples in \S\ref{sec:acf} and investigate how the size of the uncertainties in the correlation functions are impacted by masking and the available path length. We summarize our main results in the final section.

\section{Treatment of BALs} \label{sec:bals} 

\begin{figure}
    \centering
    \includegraphics[scale=0.3]{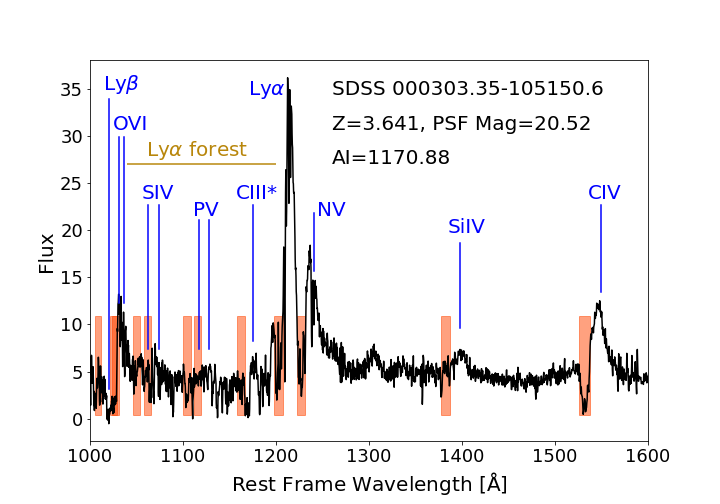}
    \includegraphics[scale=0.3]{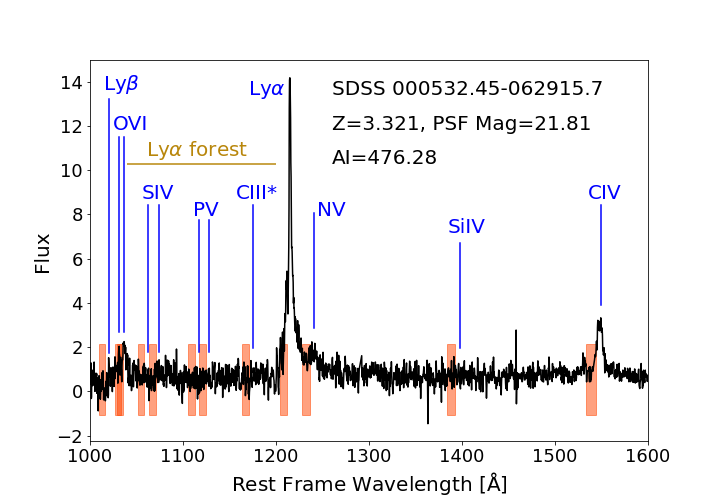}
    \includegraphics[scale=0.3]{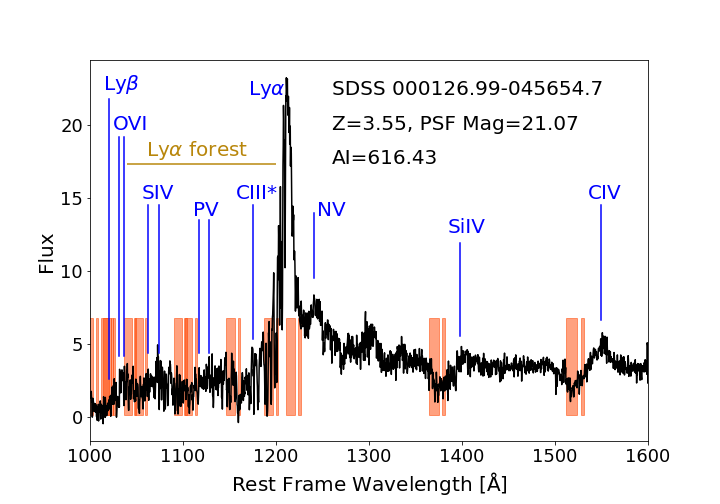}
    \caption{Three example spectra of BAL quasars that show from the $1040 - 1200$\,\AA\, region used for Ly$\alpha$ forest analysis through to the \ion{C}{iv} emission line at 1549\,\AA\,. These quasars were identified as BALs by \citet{Guo_2019}, who used a convolutional neural network classifier to search for BAL features associated with the \ion{C}{iv} emission. The wavelength range affected by \ion{C}{iv} absorption is shaded in red. We mark the equivalent wavelengths of blueshifted absorption associated with the \ion{N}{v}, Ly$\alpha$, \ion{C}{iii*}, \ion{P}{v}, \ion{S}{iv}, \ion{O}{vi}, and Ly$\beta$ emission lines. In some cases the BAL features associated with these other features are difficult to detect due to the \ion{H}{i} absorption that comprises the forest. The three examples include the spectra of an above average AI value ({\it top}), a below average AI value ({\it middle}), and one with a sufficiently large blueshift that the Ly$\alpha$ BAL feature appears below 1200\AA\ and consequently falls in the region used for analysis ({\it bottom}). "Z" refers to the redshift and "PSF Mag" is the g-band point spread function magnitude of the quasar. Flux units are $10^{-17}\rm{ergs}\:\rm{s}^{-1}\rm{cm}^{-2}\text{\AA}^{-1}$}
    \label{fig:BAL_example}
\end{figure}

The absorption features of BALs vary in width, depth, and blueshift. Accurate masking of these features requires the detection of each trough and measurement of its velocity range with respect to the emission line. In this section, we describe how BALs are identified, classified, and measured. We also describe the catalogs that contain the QSO and BAL information that were used in this study.

The classical definition of a BAL originates with the work of \citet{Weymann_1991}, who introduced the balnicity index (BI) to characterize these objects. The measurement of BI associated with \ion{C}{iv} is based on the spectrum between the \ion{C}{iv} and \ion{Si}{iv} emission lines and is defined as
\begin{equation}\label{eq:BI}
    BI = - \int_{25000}^{3000}[1-f(v)/0.9]C dv ~\rm{,}
\end{equation}
\noindent where the velocity $v$ is relative to the systemic redshift of the corresponding emission feature. This calculation identifies a BAL absorption trough if the observed flux $f(v)$ is at least 10\% below an estimate of the unabsorbed continuum, and requires that the velocity width of the absorption must be greater than 2000 km s$^{-1}$. The variable $C$ has a value of zero until both of these conditions are met, and then is set to one.  The unabsorbed continuum level is estimated in various ways, although most commonly via a fit of components derived from a Principle Component Analysis (PCA) \citep{Suzuki_2005} of a large sample of non-BAL quasars. 

The BI calculation defined by Equation~\ref{eq:BI} will ignore absorption less blueshifted than 3000 km s$^{-1}$, as well as BALs absorption features narrower than 2000 km s$^{-1}$ over the entire velocity range. To identify ouflows that would be missed by these criteria, \citet{Hall_2002} expanded on the BI definition with the introduction of the absorption index (AI), which is defined as
\begin{equation}\label{eq:AI}
    AI = - \int_{25000}^{0}[1-f(v)/0.9]C dv ~\rm{.}
\end{equation}
\noindent The AI calculation sets a minimum velocity width of 450 km s$^{-1}$ in order for $C$ to be set to one, as well as extends to blueshifts as small as the systemic velocity. The lower width was chosen to still exclude absorption due to galaxies and other structures along the line of sight, which are unlikely to exceed a width of 450 km s$^{-1}$. The AI criterion consequently identifies more BALs than the BI criterion for two reasons. First, it identifies BALs over a wider blueshift range, and second that it identifies narrower BAL features. 

Equations~\ref{eq:BI} and ~\ref{eq:AI} show that both the width and the depth of the absorption contributes to the AI and BI values. Stronger BALs, with higher AI and BI values, have more prominent absorption blueward of \ion{C}{iv}, and can be easily identified visually (Fig. ~\ref{fig:BAL_example}, top). A weaker BAL may have less obvious absorption (Fig. ~\ref{fig:BAL_example}, middle), and may be misidentified as a non-BAL in visual inspection. Yet these weaker absorption features can still contribute to errors in the continuum fitting process and the estimate of the absorption in the Ly$\alpha$ forest. 

Additionally, the AI and BI values do not take into account the blueshift of the absorption features. An absorption feature associated with Ly$\alpha$ and even \ion{N}{v} can appear in the forest region if it is blushifted below 1200\AA\ (Fig. ~\ref{fig:BAL_example}, bottom panel). The bottom panel of Figure ~\ref{fig:BAL_example} also shows an example of a BAL with multiple absorption troughs. AI and BI values are calculated for all absorption seen in a BAL quasar, ignoring gaps or multiple features. The AI or BI value alone therefore does not reflect the velocity ranges of the features. 

\subsection{BAL Catalogs} \label{sec:balcat}

The twelfth data release of SDSS \citep{Paris_2017} included measurements of both AI and BI for every quasar in the catalog. BALs were identified through both an automated search and an independent visual inspection. The automated search looked only at quasars with $z \geq 1.57$, where the entire \ion{C}{iv} spectral region could be detected by SDSS. The automated search classified 5.5\% of the entire quasar catalog as $BI > 0$, and 21.7\% as $AI > 0$. A visual inspection of the entire quasar catalog identified 9.9\% of all quasars with $z > 1.57$ as BALs. Quasars identified as BALs from the visual inspection were removed from the DR12 Ly$\alpha$ correlation analyses \citep{dumasdesbourboux17,Bautista_2017}. 

With the larger catalog size of DR14, \citet{Paris_2018}, only identified BALs by an automatic process on quasars with $z > 1.57$ and only employed the BI criterion, and not the AI criterion. About 7.4\% of quasars in this range had $BI > 0$ and were removed from the Ly$\alpha$ correlation function analyses \citep{Blomqvist_2019, de_Sainte_Agathe_2019}.

\citet{Guo_2019} developed a convolutional neural network (CNN) to identify BALs based on the visual classifications from DR12, and applied this classifier to the DR14 QSO sample. The classifier was based on the the presence of blueshifted absorption associated with the \ion{C}{iv} line, and therefore was only applied to quasars with redshifts between $1.57 < z < 5.56$. The CNN assigned a BAL probability to each quasar, and \citet{Guo_2019} identified quasars with a BAL probability $> 0.5$ as BAL quasars. This criterion identified 16.8\% of the quasars in this redshift range as BALs, in good agreement with previous measurements of the BAL fraction in DR12 and DR14. 

We used this DR14 BAL catalog in our study because it also includes measurements of AI, BI, the number of absorption troughs that satisfy each criterion, and especially because it includes the minimum and maximum velocity of each absorption trough relative to the systemic redshift. We use these velocity limits to infer the wavelength range that may be impacted by BAL features associated with other lines in the spectrum of each BAL, most notably those absorption features that overlap the wavelength range for Ly$\alpha$ forest analysis. We note that the \citet{Paris_2018} DR14 QSO catalog also identifies BAL quasars, but it only identifies them based on the BI criterion, and it does not include the velocity limits for each absorption trough that are required to mask these features. In addition to the velocity limits, we wished to mask both the stronger absorption features that met the BI criterion and the weaker features that satisfied the AI criterion.    

\subsection{BAL masking} \label{sec:balmask}
BALs are most commonly identified based on absorption blueward of \ion{C}{iv} $\lambda1549$. This is mainly because nearly all BALs exhibit absorption associated with \ion{C}{iv}, although it is also easier to detect BALs in this spectral region because the blue side of \ion{C}{iv} is relatively featureless. Yet BALs are important for studies of the Ly$\alpha$ forest because BALs are known to show absorption from other UV emission lines, including a number that are in or near the forest region. Figure~\ref{fig:BAL_example} shows some examples of broad absorption features.

Stacking analysis of BALs by \citet{Mas_Ribas_2019} and \citet{hamann19} show absorption associated with many lines. These include \ion{P}{v} $\lambda\lambda1118, 1128$, \ion{S}{iv} $\lambda\lambda 1062, 1074$, and \ion{C}{iii*} $\lambda 1175$ that fall in the forest continuum region. There is also absorption associated with \ion{N}{v} $\lambda 1241$ and Ly$\alpha$ $\lambda 1216$, which are at somewhat longer wavelengths than the forest region. These lines are important because absorption associated with these lines can be blueshifted into the forest continuum region by sufficiently high velocity outflows. Absorption from \ion{O}{vi} $\lambda 1037$ is also present, but falls outside the region used for Ly$\alpha$ forest analysis. 

These stacking studies, as well as studies of individual BALs at lower redshifts \citep[]{Weymann_1991,Arav_2001}, also show that the blueshifted velocities of the absorption troughs associated with different spectral lines are similar, which supports the hypothesis that the lines have a common origin in highly-ionized, outflowing gas. This is important, as it means that the velocity limits of the absorption troughs may be measured from the absorption troughs associated with \ion{C}{iv}, and then the corresponding locations of the absorption associated with the other lines may be masked out if the troughs fall in the forest continuum. 

Another result of these stacking studies was that the strongest BALs of DR12 studied in \citet{hamann19} appear to have significant continuum differences in the Ly$\alpha$ forest region relative to non-BAL QSOs. If there are significant differences between the unabsorbed continuum of BALs and non-BALs, then this could lead to systematic differences in the Ly$\alpha$ forest analysis. This is because the analysis software applied for BOSS and eBOSS fits a single, mean continuum shape multiplied by a first-order polynomial to each QSO \citep{Busca_2013, delubac15,Bautista_2017}. If the mean continuum based on all quasars is not a good match to BALs, then the measured Ly$\alpha$ absorption for the BALs would have larger systematic errors. 

To investigate potential continuum differences, as well as the broader impact of BALs on cosmological analysis, we mask the expected locations of all potential BAL features associated with Ly$\alpha$, \ion{N}{v}, \ion{C}{iii*}, \ion{S}{iv}, and \ion{P}{v}. Our procedure is to take the velocity limits of the absorption trough(s) based on the \ion{C}{iv} region and mask out the corresponding velocity range associated with each of these other features. We mask these regions regardless of whether or not absorption is apparent through visual inspection, as in most cases we expect the features to be too weak to separate from the IGM absorption in the forest region. While this may mask out some pathlength that is unaffected by BALs, for most BALs it is a relatively small fraction of the total pathlength, and in any case there is still a net increase in pathlength by including these BALs in the analysis. 

We used the \texttt{picca}\footnote{Available at \texttt{https://github.com/igmhub/picca}} package to develop and test the BAL mask. This software was developed for the SDSS Ly$\alpha$ forest studies of the auto- and cross-correlation \citep[e.g.][]{du_Mas_des_Bourboux_2020}, and will be used by the DESI project. The masking options include to mask features identified by either the AI or BI criteria. Within \texttt{picca}, the BAL absorption wavelengths are masked in a similar manner to other features, such as sky lines and damped Lyman-alpha systems.

\begin{table}
    \centering
\begin{tabular}{|c|c|c|}
    \hline
     Percent of BALs & AI Range & Total Number of Quasars \\
     \hline
     0\%   & AI = 0      & 267115 \\ 
     25\%  & 0 < AI $\leq$ 249.8   & 13437 \\
     50\%  & 249.8 < AI $\leq$ 839.0 & 13433 \\
     75\%  & 839.0 < AI $\leq$ 2221.6  & 13438 \\
     100\% & AI > 2221  & 13437 \\
     \hline
\end{tabular}
    \caption{Number of quasars in each AI quartile.}
    \label{tab:Quartiles}
\end{table}

We constructed five QSO subsamples to investigate the continuum shape of BALs in the forest region. Starting with the BAL catalog of \citet{Guo_2019}, we split QSOs with \texttt{AI\_CIV}$ > 0$ into four quartiles based on the value of \texttt{AI\_CIV}, which we adopt as a proxy for the amount of absorption. Each quartile includes $\sim13400$ BAL quasars with $1.57 < z < 5.56$, with ranges $0 < AI \leq 249.8$, $249.8 < AI \leq 839.0$, $839.0 < AI \leq 2221.6$, and $AI > 2221.6$. In addition, there are 267115 non-BAL ($AI = 0$) quasars in this same redshift range. This sample is somewhat smaller than the non-BAL sample used by \citet{de_Sainte_Agathe_2019}, as that study calculated the correlation function with QSOs that satisfied the less-restrictive criterion $BI = 0$. The quartile samples are summarized in Table~\ref{tab:Quartiles}

\section{QSO Continuum} \label{sec:cont}

We investigated the continuum shape of BALs with the \texttt{picca} package. This package assumes the continuum $C_q (\lambda)$ in the forest region for any quasar can be described as the product of a universal mean continuum for all quasars $C(\lambda_{RF})$ and a linear function of ${\rm log}\lambda$ that accounts for the diversity of quasar spectra. The relation between the two quantities is 
\begin{equation}
    C_q(\lambda) \bar{F} = C(\lambda_{RF}) (a_q + b_q {\rm log} \lambda) ~\rm{.}
    \label{eqn:c_q}
\end{equation}
The mean continuum $C(\lambda_{RF})$ is normalized to unity and both this function and the quantities $(a_q,b_q)$ for each quasar are calculated via a maximum likelihood approach that simultaneously calculates the mean absorption \citep[for more details see][]{Bautista_2017}. $\bar{F}$ is the mean transmitted fraction.

\begin{figure}
    \centering
    \includegraphics[scale=0.35]{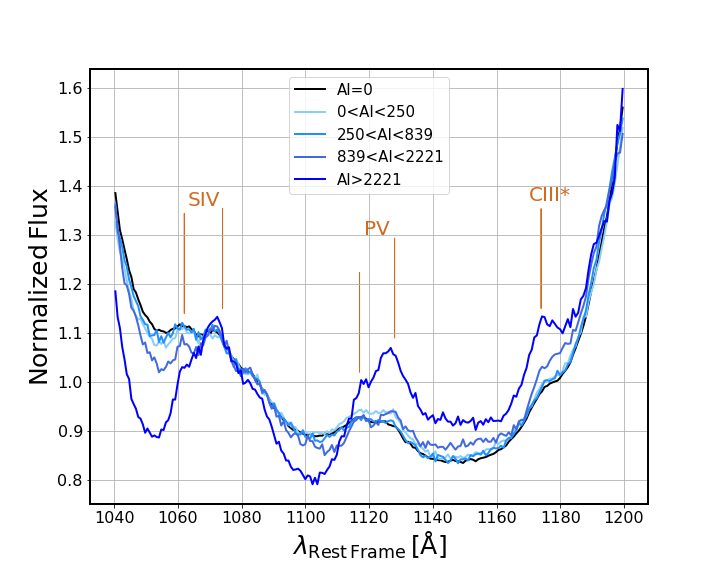}
    \caption{The continuum $\times$ mean flux of the four BAL samples from Table ~\ref{tab:Quartiles} without masking, and the AI=0 sample from DR14. Each wavelength has an average flux calculated, then the fluxes for each sample are normalized to one. The strongest quartile of BALs has a different shape than the weaker quartiles. }
    \label{fig:MeanCont_BALQuarts_NoMask.png}
\end{figure}

We computed the mean continuum for the non-BAL QSOs and for each of the four quartiles of BAL QSOs, both with and without masking the BAL features, to evaluate the effectiveness of the masking and if there are intrinsic differences in the mean continuum shape for BAL QSOs. Figure~\ref{fig:MeanCont_BALQuarts_NoMask.png} shows the mean flux of non-BALs ($AI = 0$) and the four BAL subsamples without any masking. This figure shows that the non-BAL QSOs and the first two quartiles ($AI \leq 839$) are quite similar, and therefore that there is relatively little difference in continuum shape for the weakest BALs relative to non-BAL QSOs. This was also shown by the stacking study of \citet{hamann19}. The exception is that there are some slight differences on the blue side of the \ion{S}{iv} and \ion{P}{v} lines. The third quartile ($839 < AI \leq 2222$) shows more pronounced differences, including on the blue side of the \ion{C}{iii*} line, and the fourth quartile shows a very substantial difference. These differences are largely in the region around known BAL features, which suggests they may be primarily due to the BAL features, rather than intrinsic differences in the continuum shape. 

\begin{figure}
    \centering
    \includegraphics[scale=0.35]{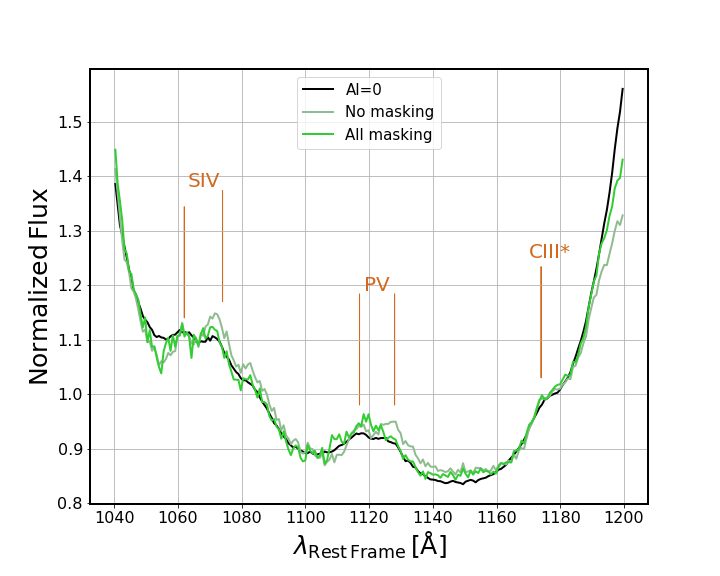}
    \caption{The continuum $\times$ mean flux of the third quartile of BALs (839<AI<2221), with different masking strategies, compared to the AI=0 sample (black).}
    \label{fig:MeanCont_AI75_MaskTest}
\end{figure}

We investigate how well masking the BAL features can explain and potentially mitigate the differences with a study of the third quartile of AI strength, as that is the lowest-AI quartile where there were clear differences in Figure~\ref{fig:MeanCont_BALQuarts_NoMask.png}. Figure~\ref{fig:MeanCont_AI75_MaskTest} shows the mean continuum for non-BALs and the third quartile continuum computed with and without masking the BAL features. The non-BAL continuum and the third quartile continuum without masking are the same as those shown in Figure~\ref{fig:MeanCont_BALQuarts_NoMask.png}. Unlike the continuum without masking, the continuum with masking is nearly identical to the non-BAL continuum. This is evidence that the BAL features are largely responsible for the apparent differences in the continuum shape for the BALs, and that masking the features produces a very similar mean continuum as the non-BAL QSOs. 

\begin{figure}
    \centering
    \includegraphics[scale=0.35]{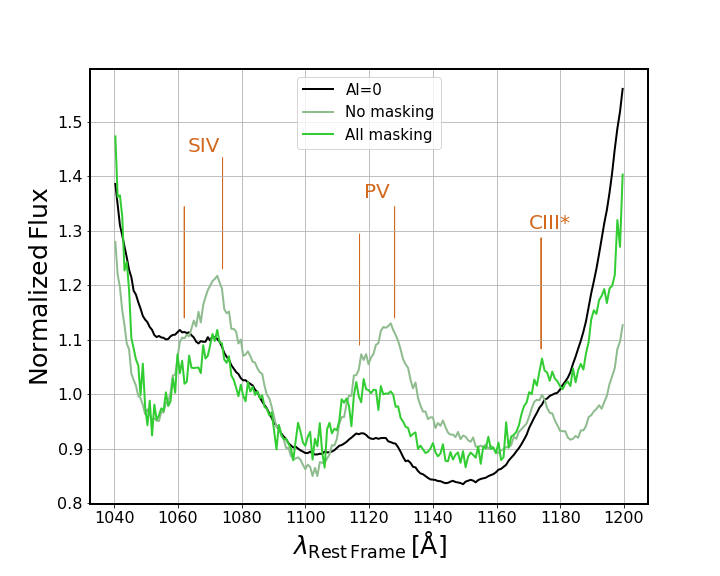}
    \caption{The continuum $\times$ mean flux of the strongest BALs. Masking the forest region lines of \ion{S}{iv}, \ion{P}{v}, and \ion{C}{iii*} results in a continuum shape more like the AI=0 shape. However, these regions are very noisy due to large amounts of masking.}
    \label{fig:MeanCont_AI100_MaskTest}
\end{figure}

We repeat this experiment for the fourth quartile ($AI > 2222$) in Figure~\ref{fig:MeanCont_AI100_MaskTest}. This figure shows that masking the BAL features does not recover the non-BAL continuum for the fourth quartile. Prominent differences are present on the blue side of the lines associated with BALs in both the masked and non-masked cases. While the BAL features are somewhat less prominent in the continuum constructed with masking, they are nevertheless still present. These differences may represent limitations of the masking procedure and/or genuine differences in the continuum shape.

\begin{figure}
    \centering
    \includegraphics[scale=0.38]{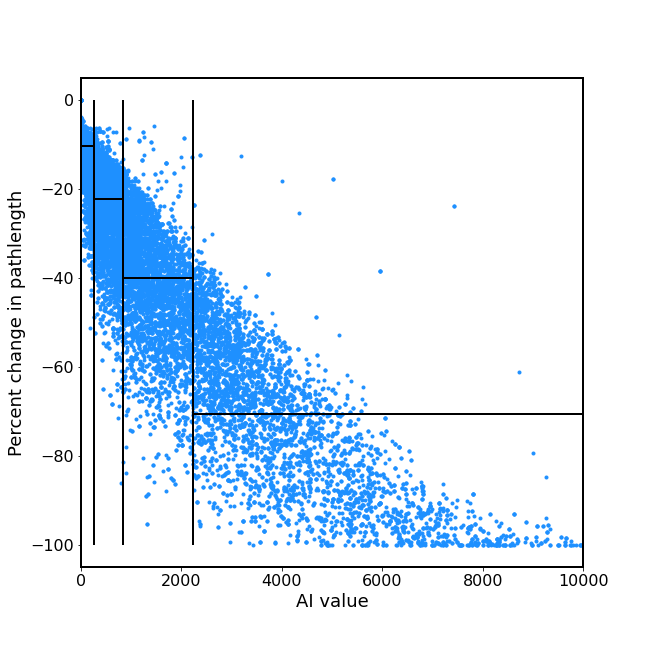}
    \caption{The percent change in pathlength for individual quasars due to masking.
    We randomly selected 5000 quasars with z > 2.57 from each AI subsample (Table ~\ref{tab:Quartiles}. The AI value is a calculation of both the width and the depth of the absorption. For masking, we only need the width.}
    \label{fig:PLchange_forAI}
\end{figure}

To explore the origin of these differences further, as well as to quantify the impact of not including the largest quartile of BALs, we computed the percent of the pathlength through the forest that is masked in each BAL QSO. BALs with a relatively narrow feature will have only a small fraction of the total pathlength masked, while those with broad features may have a very substantial fraction of the total pathlength masked. Figure~\ref{fig:PLchange_forAI} shows the percent change in pathlength vs.\ AI for a subset of 5000 BAL QSOs. There is some pathlength decrease for all BALs, as the minimum velocity extent of a BAL feature is $450\, {\rm km\, s}^{-1}$. The median pathlength varies from less than 10\% for the first quartile to about 70\% for the fourth quartile, and there are some BALs in the fourth quartile where the entire forest continuum region is masked due to overlapping BAL features. There is substantial scatter about this relationship because AI is a measure of the strength (equivalent width) of the absorption trough(s), rather than the velocity width of the trough(s). 

The median masked pathlength of 70\% for the fourth quartile indicates that there is relatively less value to including the fourth quartile of BALs in the cosmological analysis. The substantial percentage of the pathlength that is masked may also explain why the mean continuum of the fourth quartile BALs did not match the non-BAL mean continuum, even after masking. These most extreme BALs had relatively little unmasked continuum, and even none in some cases, so the determination of the mean continuum was at best noiser than for the other quartiles. There may also be yet weaker features, higher velocity outflows (greater than $25000\, {\rm km\, s}^{-1}$), and greater reddening that impact the most extreme subset of the BAL population.  

\section{Autocorrelation Function} \label{sec:acf}

In this section we investigate the impact of including BAL QSOs in the calculation of the autocorrelation function of the Ly$\alpha$ forest. We first provide an overview of the calculation details that could be impacted by continuum differences and our BAL masking procedure. This calculation closely follows the analysis of DR14 presented by \cite{de_Sainte_Agathe_2019}. We then quantify how the addition of BAL QSOs impacts the measurement of the autocorrelation function.

\subsection{Calculation} \label{sec:calc}

Ly$\alpha$ absorption due to intergalactic gas produces variations in the optical depth that are related to variations in the matter distribution along the line of sight. The autocorrelation function is calculated from variations in the transmitted flux relative to the average flux transmission at the absorber redshift $\Bar{F}(z)$. The variation $\delta_q(\lambda)$ in the transmitted flux is defined as
\begin{equation}
    \delta_{q}(\lambda) = \frac{f_{q}(\lambda)}{C_{q}(\lambda)\Bar{F}(z)} -1 ~\rm{,}
    \label{eqn:delta} 
\end{equation} 
where $f_q(\lambda)$ is the observed flux of the quasar at wavelength $\lambda$ and $C_q(\lambda)$ is the intrinsic continuum of the quasar at that wavelength. 

In practice, \texttt{picca} fits the product of the intrinsic continuum and $\bar{F}(z)$ as described in \S\ref{sec:cont}, and the spectra are split into distinct analysis pixels that each correspond to three native pixels as output by the spectroscopic pipeline. This has a minimal impact on the BAO scale and reduces the computation time. The flux in each analysis pixel is then the average of the three native pixels weighted by their inverse variance.  

The \texttt{picca} code assigns a weight $w_q(\lambda)$ to each analysis pixel based on the inverse variance. The inverse variance $\sigma_q(\lambda)$ has three contributions: the pipeline estimate of the noise variance, density fluctuations in the Ly$\alpha$ forest, and variance due to the greater diversity of real quasar spectra relative to the fit to the continuum flux $C_q(\lambda)$. This third contribution will be greater for BAL QSOs if they exhibit greater spectral diversity. 

The correlation function at each separation $r$ is the weighted average of all of the $\delta$ values at that separation.
\begin{equation}
    \xi(\mathbf{r}) = \frac{\Sigma_{i,j} w_i w_j \delta_i \delta_j}{\Sigma_{i,j} w_i w_j}
\end{equation}
where $\delta_i$ and $\delta_j$ come from pixels in different quasars whose separation is $\mathbf{r}$. This separation has two components, $r_{\perp}$ and $r_{\parallel}$. Figure 5 in \citet{de_Sainte_Agathe_2019} gives a visual reference. To simplify the calculation, the two components $r_{\perp}$ and $r_{\parallel}$ are each calculated in 4 h$^{-1}$Mpc bins over the range 0 - 200 h$^{-1}$Mpc. We consequently calculate the correlation function on a $50\times50$ element grid. Each element of the grid has a corresponding $50\times50$ element covariance matrix.

Following \citet{de_Sainte_Agathe_2019}, we separate this 2D grid into four wedges with the \texttt{baoutils} package\footnote{Available at \texttt{https://github.com/julienguy/baoutil}} and calculate the correlation function and uncertainty in each wedge as a function of $\mathbf{r}$. Each wedge covers a range $\mu = \mathbf{r}/r_{\perp} = cos(\theta)$ that range from mostly transverse to mostly parallel to the line of sight. Bins that fall along the boundaries of the wedges are subsampled onto a $7\times7$ grid. We calculate the uncertainties in each bin by propagating the relevant bins in the covariance matrix.

\subsection{Impact of Masking}

\begin{table}
    \centering
\begin{tabular}{|c|c|c|}
    \hline
     Percent of BALs & Max AI Value & Total Number of Quasars \\
     \hline
     0\%   & 0         & 267115  \\ 
     25\%  & 249.8     & 280552  \\
     50\%  & 839.0     & 293985  \\
     75\%  & 2221.6    & 307423  \\
     100\% & No Limit  & 320860  \\
     \hline
\end{tabular}
    \caption{Maximum AI value and total number of quasars in each set of autocorrelation runs.}
    \label{tab:AI_cat_range}
\end{table}

We evaluated the impact of adding masked BALs to the calculation of the correlation function by progressively adding the BAL samples described in \S\ref{sec:cont} to the $AI = 0$ sample. We performed the autocorrelation calculation five times, on the samples listed in Table~\ref{tab:AI_cat_range}. The sample with
non-BALs only ($AI=0$) is somewhat smaller than the sample used by \citet{de_Sainte_Agathe_2019}, as that study calculated the correlation function with QSOs that satisfied the less-restrictive criterion $BI = 0$. That study also used the DR14 QSO catalog \citep{Paris_2018} rather than the \citet{Guo_2019} BAL catalog that we employ, although that is a minor difference as the $BI$ measurements for these two catalogs are quite similar \citep{Guo_2019}.

\begin{figure}
    \centering
    \includegraphics[scale=0.35]{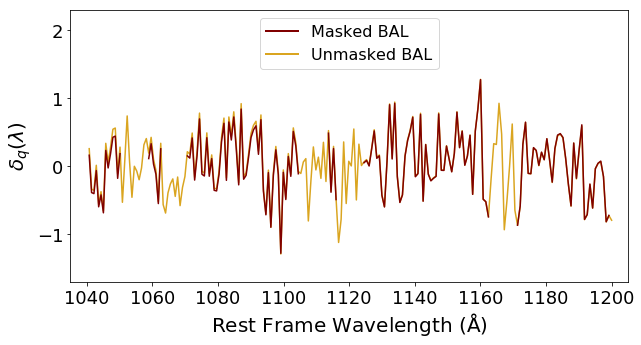}
    \caption{The effects of masking BAL features on $\delta_{q}(\lambda)$ for a BAL quasar (see Eqn. ~\ref{eqn:delta}). The light, gray line shows the observed flux of the quasar as a function of wavelength. The line labeled "Unmasked BAL" is $\delta_{q}(\lambda)$ vs.\ wavelength computed with a continuum fit to the entire forest region, while the "Masked BAL" is from a continuum fit to only pixels that do not contain potential BAL features. The two $\delta_{q}(\lambda)$ lines are slightly offset because the continuum is fit to a subset of the data when the potential BAL features are masked. 
    }
    \label{fig:MaskingDeltas}
\end{figure}

\begin{figure*}
    \centering
    \includegraphics[scale=0.45]{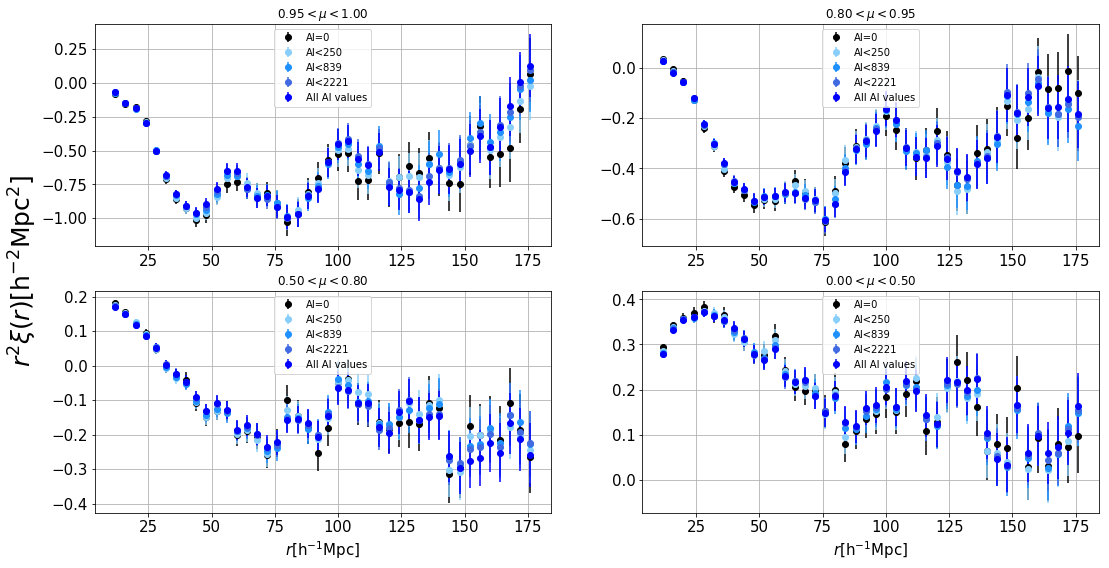}
    \caption{The 2D autocorrelation results for AI=0 quasars, AI<250, AI<839, AI<2221, and all quasars. The correlation values calculated with BALs generally remain within error of the AI = 0 sample. }
    \label{fig:Autocorr_Quarters}
\end{figure*}

First we examine the impact of masking on $\delta_{q}(\lambda)$. The calculation of $C_{q}(\lambda)\Bar{F}(z)$ is fit over the entire Ly-$\alpha$ forest analysis region of a given quasar (see Eqn. ~\ref{eqn:c_q}). We mask BAL features before $\delta_{q}(\lambda)$ is calculated, so we expect some change in the calculated deltas of a masked BAL quasar, compared to the unmasked case, such as in Figure ~\ref{fig:MaskingDeltas}. These changes are quite small, and result in small variations in the autocorrelations of Figure ~\ref{fig:Autocorr_Quarters}. 

Our autocorrelation function calculations for these five samples are shown in Figure ~\ref{fig:Autocorr_Quarters}. This figure shows that while there are some slight shifts in the autocorrelation values, these changes are largely within the uncertainties of the $AI = 0$ sample. In addition, there are no trends for larger deviations as a function of $r$ for any of the wedges. This behavior is unsurprising, as the locations of BAL features are uncorrelated with large scale structure. 

\begin{figure*}
    \centering
    \includegraphics[scale=0.45]{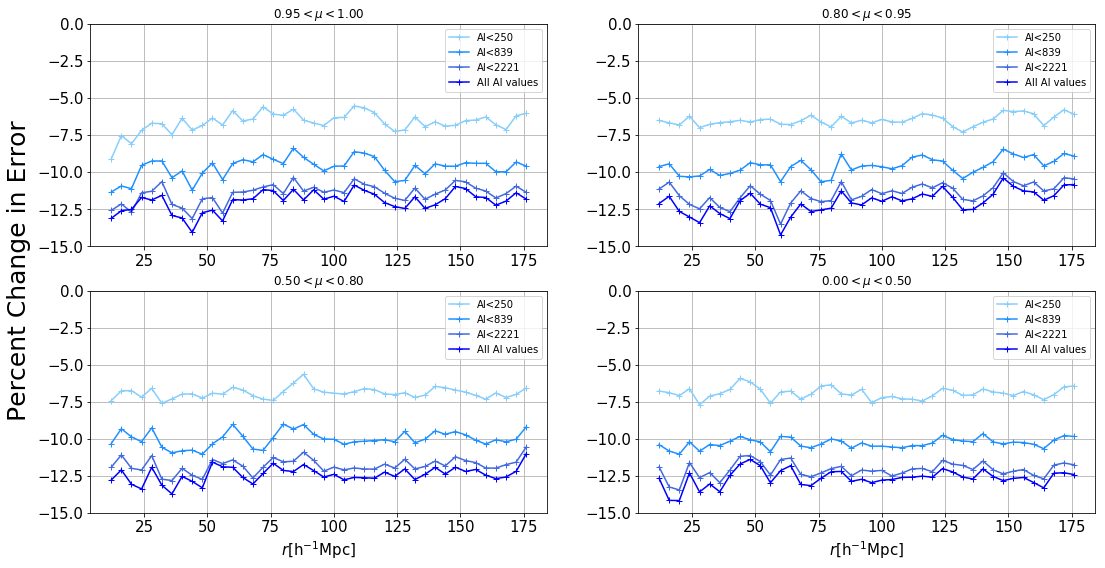}
    \caption{Percent change in 2D Autocorrelation Errors in each wedge plot. The uncertainty decreases by about 12\% when including masked BALs, compared to the AI=0 only sample. }
    \label{fig:CorrErrors_AllMasking}
\end{figure*}

The more relevant quantity is the impact on the uncertainty in the correlation function. The addition of each progressive quartile of BAL QSOs increases the total pathlength and the surface density of QSOs, so we expect a corresponding improvement in the fractional errors. This is showns in Figure ~\ref{fig:CorrErrors_AllMasking}, which expresses the percent change in the fractional errors as a function of $r$ for each subsample relative to the $AI=0$ case for the four wedges shown in Figure~\ref{fig:Autocorr_Quarters}. The overall trend is that the addition of the first quartile leads to a $\sim 7$\% decrease in the uncertainties, the addition of the first two quartiles leads to a $\sim 10$\% decrease, and the addition of the third and then the fourth quartile produce decreases of $\sim12$\% and $\sim 13$\%, respectively. There is some scatter in the percent improvement as a function of $r$ for all of these subsamples. This is expected from the variation in the number of pairs in each bin. 

\begin{figure*}
    \centering
    \includegraphics[scale=0.45]{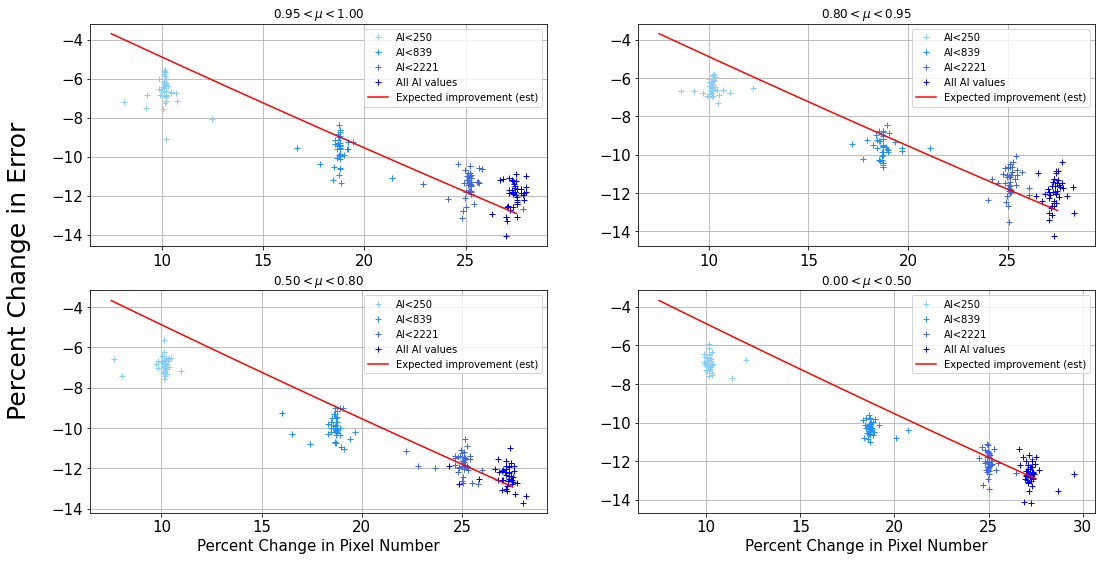}
    \caption{The percent increase in the number of pixel pairs going into each calculation of the autocorrelation point, compared to the reduction in error. The cloud of points for each sample corresponds to the bins of $r$ for that wedge of the autocorrelation function. The straight red line represents the expected decrease, if we treat the data as purely statistical and approximate the errors as $\sqrt{N}$, where $N$ is the number of pixel pairs. The errors decrease following this line, suggesting that the data we are including is similar to non-BAL data.}
    \label{fig:ErrorvsPixPairs}
\end{figure*}

There are at least two contributions that produce the progressively more modest gains with the inclusion of each successive quartile. One is that the fractional increase in the sample size is progressively smaller, as we are adding a constant number of QSOs to a progressively larger sample. The second is that while each subsample adds the same number of QSOs, the unmasked pathlength of each subsample is progressively less. We investigate these two effects with Figure~\ref{fig:ErrorvsPixPairs}, which shows the percent change in the fractional error after the inclusion of each quartile as a function of the percent change in the total number of pixel pairs. This figure shows that there is an approximately linear relationship between the percent decrease in the fractional error and the percent increase in the number of pixel pairs. 

If the fractional errors are purely statistical, then we expect the percent decrease in the fractional errors will scale as $\sqrt{N}$, where $N$ is the number of pixel pairs used in each autocorrelation calculation. This relationship is shown as the solid line in Figure~\ref{fig:ErrorvsPixPairs}. While the data generally follow this relationship, the improvements from the first two quartiles are somewhat better than this expectation, while the improvement from the addition of the last quartile is somewhat worse. This difference is likely because the weights described in \S\ref{sec:calc} depend on the signal-to-noise ratio (SNR) of the data, as it is easier to detect weaker BALs in higher SNR spectra. Another factor is that this assumption of purely statistical errors is not correct, as there are correlations between the pixels that are used to estimate the errors. 

\begin{figure}
    \centering
    \includegraphics[scale=0.45]{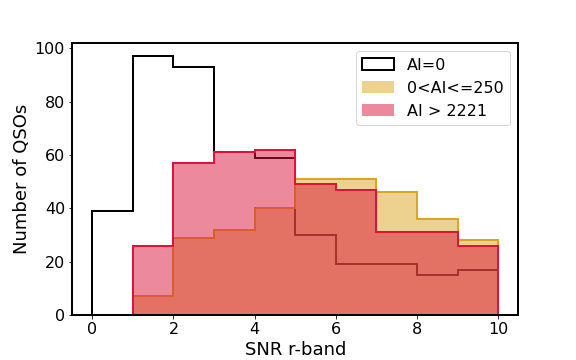}
    \caption{Distribution of $r-$band SNR for non-BAL ($AI=0$) quasars and the lowest and highest quartile of $AI$ value for BAL quasars. The SNR distribution for the quasars with $AI=0$ is much lower than the two BAL samples. The SNR distribution of the lowest AI quartile ($0 < AI \leq 250$) has somewhat higher SNR than the highest AI quartile ($AI < 2221$). }
    \label{fig:balsnr}
\end{figure}

We tested this hypothesis with the median SNR measurements computed by the SDSS spectroscopic pipeline for each of the SDSS photometric bands. We used the $r-$band values as this region should be relatively unaffected by BAL and IGM absorption features. We randomly selected a subset of 500 quasars with $2.1 < z < 2.3$ in each of three categories: non-BALs with $AI = 0$; BALs in the first quartile of AI strength; and BALs in the fourth quartile of AI strength. Both BAL samples have a median SNR that is approximately a factor of two larger than the non-BALs with $AI = 0$, with somewhat higher SNR for the quasars with the lowest quartile of $AI$. This result strongly supports our explanation for why we see greater improvement in the fractional errors when we add the first quartile of BALs. The distributions of these three subsamples are shown in Figure~\ref{fig:balsnr}. 

The higher SNR of the BAL quasars makes them more valuable in a per pixel sense than non-BAL quasars, although we note that the lower SNR of the $AI=0$ sample also means that some of these quasars are  BALs that were simply not identified in the lower-quality data. Fortunately these weaker features also have less impact on the continuum fitting and correlation function. The BALs with larger $AI$ values are easier to detect in lower SNR spectra, although the median SNR of this sample is also about a factor of two greater than the $AI = 0$ sample. This suggests that there may be some more significant BAL quasars in the $AI = 0$ sample as well. In spite of their somewhat higher SNR ratio relative to non-BALs, the percent improvement in the correlation function uncertainties for the largest quartile in AI strength is approximately in agreement with the expectations of Poisson statistics. This may be due to other effects, most notably greater uncertainty in continuum fitting due to the substantial fraction of their wavelength range that is masked.  

\section{Summary}

In this study we have investigated how to include BAL quasars in measurements of the Ly$\alpha$ forest autocorrelation function, which is the primary tool for distance measurements with quasars above $z > 2$. BAL features are typically observed in the spectra of 10--20\% of all quasars selected for large-scale spectroscopic surveys, and the exclusion of BAL quasars in the analysis of the previous generation of surveys has led to a corresponding decrease in the total forest path length employed for cosmological studies. 

As BAL features are due to highly-ionized, blueshifted gas clouds, the absorption features are typically present over the same velocity range for all ionic species that are present in absorption. The BAL catalog of \citet{Guo_2019} identified \ion{c}{iv} BAL troughs associated with 16.8\% of the quasars in the SDSS DR14 quasar catalog. The parent sample for that percentage was all quasars whose spectra included the wavelength region on the blue side of \ion{C}{IV} ($1.57 < z < 5.6$). That catalog includes the velocity limits of each trough, and we have used these limits to mask the corresponding wavelength range for \ion{O}{VI}, \ion{S}{IV}, \ion{C}{III*}, Ly$\alpha$, and \ion{N}{V}. While not all BAL quasars exhibit absorption in all of these species, these features are apparent in stacked spectral studies \citep{hamann19,Mas_Ribas_2019}. As we cannot distinguish BAL features from the forest in most cases to be confident they are not present, it is consequently more conservative to mask them.

One metric we used to evaluate the performance of masking was to reconstruct the mean quasar continuum in the forest region based on known BALs with the troughs masked, and to compare that continuum to the reconstructed mean continuum of non-BAL quasars, with a calculated Absorption Index (AI) of $0$. The continuum shape is a useful metric because if BAL quasars have different continuum shapes than non-BAL quasars, this would add systematic errors to the cosmological analysis. Alternatively, if there are additional, unmasked BAL features, this could manifest as differences in the continuum shape between the BAL and non-BAL samples. We split the BAL sample into quartiles based on AI value and constructed the continuum with \texttt{picca} for each quartile, as well as for the larger sample of non-BALs. This study showed that the reconstructed continuum shape for the first three lowest AI quartiles ($AI < 2221$) was consistent with the continuum constructed from non-BAL quasars, while the fourth quartile with the largest AI values exhibited significant differences in the vicinity of the trough locations.

The second metric was to investigate the impact of adding BAL quasars on the fractional errors in the autocorrelation function. For this analysis, we computed the autocorrelation function with the non-BAL sample, and then calculated the decrease in the fractional errors after successively adding each quartile of AI. The result was that the addition of each of the first three quartiles leads to a notable decrease in the fractional errors in the autocorrelation function, while the addition of the fourth quartile led to a minimal improvement.

For each quartile, we also calculated the fractional increase in the number of forest pixel pairs used in the correlation function calculation. We found that the decrease in the fractional errors after the addition of the first two quartiles was actually greater than expected from Poisson statistics alone. This is because the SNR of the spectra of these quartiles is greater than that of non-BALs, and therefore the BAL forest pixels have greater weight in the autocorrelation function calculation. One caveat is that the greater SNR of the BAL sample is because it is easier to detect BALs in higher SNR spectra, so the non-BAL sample will consequently be impacted by unidentified and therefore unmasked BAL troughs. The gain with the addition of the third and fourth quartiles is much less. This is likely because it is easier to identify stronger BALs in lower SNR data. It may also be the case that the continuum fit is lower accuracy because more of the continuum of each BAL is masked. 

The addition of BAL quasars will benefit both the re-analysis of current datasets and the analysis of future experiments such as the Dark Energy Spectroscopic Instrument \citep{desi16b}. Our work has shown that the inclusion of the three lowest AI quartiles of BALs ($AI < 2221$) in eBOSS DR14 decreases the fractional error in the Ly$\alpha$ autocorrelation function by approximately 12\% due to the improved sampling of the density field. This gain is equivalent to extending the survey area by nearly 25\%, which would extend the 5-year DESI survey by over a year (for Ly$\alpha$ quasars). We also find there is minimal gain with the addition of the strongest quartile of BALs with respect to AI value. The main reason for this minimal gain is that most of the forest pathlength of these quasars is masked. Another potential contribution is that additional, yet weaker BAL features may affect their spectra. As the value of BAL quasars for correlation studies is most closely connected to the total unmasked pathlength, we recommend the use of the velocity extent of the BAL features as an additional criterion for whether or not to include a BAL quasar in future analysis. 

\section*{Acknowledgements}

LE and PM acknowledge support from the United States Department of Energy, Office of High Energy Physics under Award Number DE-SC-0011726, and PM is grateful for support from the Radcliffe Institute for Advanced Study at Harvard University while some of this work was completed. We also acknowledge helpful discussions with the DESI Ly$\alpha$ forest working group, and with Victoria de Sante Agathe and Julien Guy at the start of this project. 

AFR is supported by MICINN with a Ram\'on y Cajal contract (RYC-2018-025210). IFAE is partially funded by the CERCA program of the Generalitat de Catalunya.

This material is based upon work supported by the National Science Foundation Graduate Research Fellowship Program under Grant No. DGE-1343012. Any opinions, findings, and conclusions or recommendations expressed in this material are those of the authors and do not necessarily reflect the views of the National Science Foundation.

This research used resources of the National Energy Research Scientific Computing Center (NERSC), a U.S. Department of Energy Office of Science User Facility located at Lawrence Berkeley National Laboratory, operated under Contract No. DE-AC02-05CH11231.

Funding for the Sloan Digital Sky Survey IV has been provided by the Alfred P. Sloan Foundation, the U.S. Department of Energy Office of Science, and the Participating Institutions. SDSS acknowledges support and resources from the Center for High-Performance Computing at the University of Utah. The SDSS web site is www.sdss.org.

SDSS is managed by the Astrophysical Research Consortium for the Participating Institutions of the SDSS Collaboration including the Brazilian Participation Group, the Carnegie Institution for Science, Carnegie Mellon University, Center for Astrophysics | Harvard \& Smithsonian (CfA), the Chilean Participation Group, the French Participation Group, Instituto de Astrofísica de Canarias, The Johns Hopkins University, Kavli Institute for the Physics and Mathematics of the Universe (IPMU) / University of Tokyo, the Korean Participation Group, Lawrence Berkeley National Laboratory, Leibniz Institut für Astrophysik Potsdam (AIP), Max-Planck-Institut für Astronomie (MPIA Heidelberg), Max-Planck-Institut für Astrophysik (MPA Garching), Max-Planck-Institut für Extraterrestrische Physik (MPE), National Astronomical Observatories of China, New Mexico State University, New York University, University of Notre Dame, Observatório Nacional / MCTI, The Ohio State University, Pennsylvania State University, Shanghai Astronomical Observatory, United Kingdom Participation Group, Universidad Nacional Autónoma de México, University of Arizona, University of Colorado Boulder, University of Oxford, University of Portsmouth, University of Utah, University of Virginia, University of Washington, University of Wisconsin, Vanderbilt University, and Yale University.

\section*{Data Availability}
The spectra are available from the SDSS Science Archive Server (https://data.sdss.org/sas/dr14/eboss/).

% \printbibliography
\bibliographystyle{mnras}
\bibliography{BALMaskingBib}

% Don't change these lines
\bsp	% typesetting comment
\label{lastpage}
\end{document}